# TEST RESULTS OF TESLA-STYLE CRYOMODULES AT FERMILAB *

E. Harms[#], K. Carlson, B. Chase, D. Crawford, E. Cullerton, D. Edstrom, A. Hocker, M. Kucera, J. Leibfritz, O. Nezhevenko, D. Nicklaus, Y. Pischalnikov, P. Prieto, J. Reid, W. Schappert, P. Varghese, Fermilab, Batavia, IL 60510 USA

*Abstract*

Commissioning and operation of the first Tesla-style Cryomodule (CM-1) at Fermilab was concluded in recent months. A second Tesla Type III+ module, RFCA002, will be replacing it. CM-1 is the first 8-cavity ILC style cryomodule to be built at Fermilab and also the first accelerating cryomodule of the Advanced Superconducting Test Accelerator (ASTA). We report on the operating results of both of these cryomodules.

## INTRODUCTION

An electron Linac based on Superconducting RF technology has been proposed and is under construction at Fermilab. This Advanced Superconducting Test Accelerator (ASTA) has as its main components a Cs-Te 1-1/2 cell Photoinjector gun, two Capture cavities, and Tesla–style SRF cryomodules as well as low (40 MeV) and high energy user beam lines. This facility has been described previously [1]. It currently serves as the only test bed for multi-cavity cryomodules at Fermilab until a dedicated facility, the Cryomodule Test Facility (CMTF), now under construction is completed.

The setup for testing cryomodules at ASTA contains all of the necessary infrastructure including:

- 5MW 1.3 GHz klystron capable of pulsed operation at 10 Hz
- Cryogenics plant capable of providing 120 Watts of refrigeration at 2 Kelvin
- Interlock/protection systems
- Low Level RF (Feedforward & Feedback) system
- Vacuum system
- Controls integrated to Fermilab's Accelerator Controls system, ACNET.

## CRYOMODULE 1

CM-1 was the first Tesla Type III Cryomodule to be placed into operation in the United States and the first multi-cavity cryomodule of any type to become operational at Fermilab. It was provided to Fermilab as a 'kit' from DESY in exchange for a 4-cavity 3.9 GHz module, ACC39, which was designed and assembled at Fermilab and is now in operation at DESY's FLASH facility. Table 1 highlights the major steps in bringing CM-1 into operation and completing the test program.

The test plan for CM-1 has been described previously as has been the performance characteristics of individual cavities [2]. Typical operating conditions were:

- 5 Hz repetition rate
- 700 μs fill time
- 500 μs flattop
- 2 Kelvin / 23.4 Torr temperature
- 3.9 MW peak Klystron power.

Table 1: CM-1 Commissioning Milestones

| Milestone | Date |
|---|---|
| Cryomodule moved into final position and aligned | 22 January 2010 |
| 5 MW RF/Klystron commissioning | June - July 2010 |
| Warm coupler conditioning - one cavity at a time (4 - 14 days/cavity) | August – October 2010 |
| Cool down from room temperature to 4 then 2 Kelvin | 17 – 22 November 2010 |
| Individual Cavity cold conditioning and evaluation | December 2010 – June 2011 |
| Installation of Waveguide Distribution system and Water System upgrade | 13 June - 5 July 2011 |
| First powering of the entire module | 6 July 2011 |
| Long Pulse (9ms) tests | December 2011 |
| Thermal cycling and lN$_2$ leak repair | January - February 2012 |
| Cease operation/Removal | March 2012 |

Figure 2 provides a comparison of peak cavity gradients during qualification tests at DESY as well as during CM-1 full module operation while Table 2 summarizes individual cavity performance in CM-1. On average, the cavities' peak gradients in CM-1 are 85% of that measured previously.

*Module Evaluation*

The previous report [2] on CM-1 indicated some outstanding measurements to be performed including

- Demonstration of Lorentz Force Detuning Compensation on the entire module,
- Additional processing to attempt to improve the performance of deficient cavities,

- Possible thermal cycling to improve performance/uncover additional field emitters,
- Readjustment of VTO's to better match cavity pair outputs.

In large measure this was all carried out as was a Dynamic Heat Load measurement to determine the overall cryomodule $Q_0$ vs E.

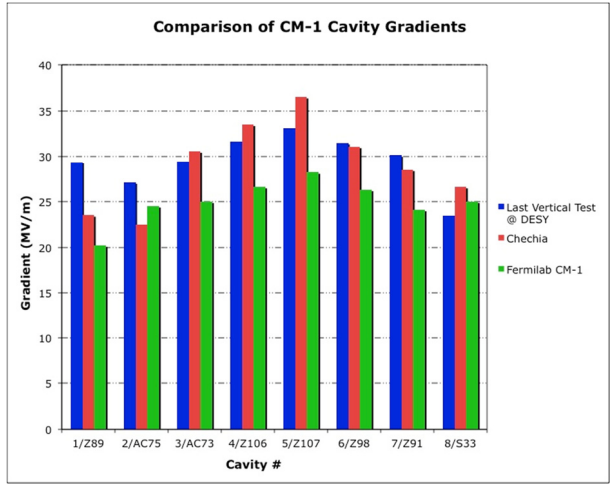

Figure 1: Comparison of cavity peak gradients – DESY single cavity vertical (blue), horizontal/Chechia (red), and the complete CM-1 at Fermilab (green).

### Lorentz Force Detuning Compensation

Adaptive Lorentz Force Detuning Compensation (LFDC) measurements made on single cavities are documented elsewhere [3] prior to when all cavities were simultaneously powered and corrected. In this state gradients in the cavities ranged between approximately 8 and 27 MV/m. Without compensation, the highest gradient cavities would detune by up to 400 Hz during the flattop. With adaptive compensation on the detuning was negligible in all the cavities except for Cavity 1 which detuned by approximately 10 Hz during the RF pulse. Even for the cavities operating at the highest gradients, Lorentz force detuning could be compensated using relatively modest voltages to drive the piezoelectric tuners, approximately ±20V, and the waveforms did not exhibit fast transients that could potentially damage them.

### Low Level RF

Development work on the Low Level RF system (LLRF) was a constant throughout the running period. By the end of the run it was possible to control the RF amplitude and phase over 50 pulses to a RMS magnitude error of $2.5 \times 10^{-4}$% and RMS phase error of 0.005° as measured in-loop.

### 9 Millisecond Test

Once the cryomodule was well characterized a 'long pulse' study in support of an R&D program for Fermilab's proposed Project-X SRF Linac was performed. This proof-of-principle study entailed operating the two best-performing cavities (5 & 6) at a pulse length of up to 9 milliseconds and determining the LLRF and resonance control capabilities at various input powers and $Q_L$ values. Operating parameters in this mode were:

- RF power limitations: 80 kW; 100 kW, 120 kW per two cavities,
- Use of the 'CC2' klystron capable of producing long pulses,
- External $Q_L$: $3 \times 10^6$; $6 \times 10^6$; $1 \times 10^7$,
- Gradient: 15MV/m; 20 MV/m; 25 MV/m,
- Pulse Width: up to 9 ms.

Table 2: Individual Cavity Performance Characteristics

| Cavity | Peak $E_{acc}$ (MV/m) | Estimated maximum $Q_0$ (E09) | Limitation/ Comments |
|---|---|---|---|
| 1/Z89 | 20.2 | 11 | 'Soft' quench |
| 2/AC75 | 22.5 | 12 | Quench |
| 3/AC73 | 23.2 | 0.43 | 'Soft' quench |
| 4/Z106 | 24* | 2.3 | *RF-limited |
| 5/Z107 | 28.2 | 39 | Quench |
| 6/Z98 | 24.5 | 5.1 | Quench |
| 7/Z91 | 22.3 | 4.7 | 'Soft' quench |
| 8/S33 | 25 | 18 | Tuner motor malfunction |

This study proved to be a good first pass test. LFDC at the nominal parameters, $Q_L = 1 \times 10^7$ and 25 MV/m, was demonstrated. LLRF feedback worked with phase stability good to ± 4°. The total range of detuning was found to be of order ±10 Hz peak-to-peak with comparable contributions from residual LFD and microphonics. The measured microphonics level was in range 2-4 Hz, similar to what was measured for short (1ms) pulses. Areas requiring further work were uncovered. Input power limitation for the low $Q_L$ case (~3 $\times 10^6$) limited the peak gradients to less than 20 MV/m. The system also proved to be less stable under dynamic conditions. Nearly constant attention was required under these conditions to maintain stable operation especially when adjusting the power.

### Thermal Cycling

Additional processing of lower performing cavities did not noticeably alter their performance. The module was warmed up to room temperature and cooled also with no significant change in performance. During cooldown a leak on the liquid nitrogen circuit appeared. In situ investigation revealed a failed feed-through on an in-line temperature sensor which was subsequently capped off and replaced with an externally mounted one. This experience proved to be valuable in both making diagnosis and repair without removing the cryomodule.

*Summary of CM-1*

Although issues with a tuner motor prevented all eight cavities from operating on resonance together and a subset of cavities exhibited sub-par performance, CM-1 did operate for an extended time stably and near the peak achievable gradients with both LLRF feedback enabled and LFDC. Valuable experience was gained and a larger cadre of staff was exposed to this type of operation.

As appropriate, components such as low performing cavities, tuner motors, and the like will be scrutinized and subjected to further testing once CM-1 is disassembled.

## RFCA002

The second cryomodule planned for ASTA was delivered in March 2012 and installation work begun immediately. This device known as RFCA002 or CM-2 is also a Tesla Type III+ device. All cavities were fabricated by industry. Vertical testing was carried out at both Jefferson Lab and Fermilab and horizontal tests of the dressed cavities at Fermilab's Horizontal Test Stand. All facets of testing and assembly have been under the supervision of Fermilab staff. Since CM-2 is being installed in place of CM-1 the necessary infrastructure is already in place.

Expectations for the cryomodule are high as individual cavity testing results for these cavities met or exceeded the ILC design specification of 31.5 MV/m. These results are described previously [4]. Figure 3 summarizes the performance.

A final leak check of the 2-phase helium circuit gave indications of a leak, so the module has been returned to its assembly building, partially disassembled, and attempts to identify the source of the leak are in progress. The current expectation is for the cryomodule to be returned to ASTA late in 2012 with full warm and cold checkout completed in the first half of 2013.

## LESSONS LEARNED

Operating experience with CM-1 in recent months is proving valuable in anticipation of bringing CM-2 and other cryomodules destined for this facility on-line in the near future as well as in preparation for possible future SRF facilities such as Project-X. Already the suite of instrumentation has been reviewed and modified for RFCA002 based on CM-1 experience. The sequence of commissioning steps has been reviewed and modified in light of the initial operating experience and will no doubt lead to faster and more efficient start-up of future modules.

Tuner motor issues with CM-1 led to a review of the motor controller circuitry, which led to modifications realizing a more robust and fail-safe system.

The Controls user interface system is also under regular updating as more experience is gained.

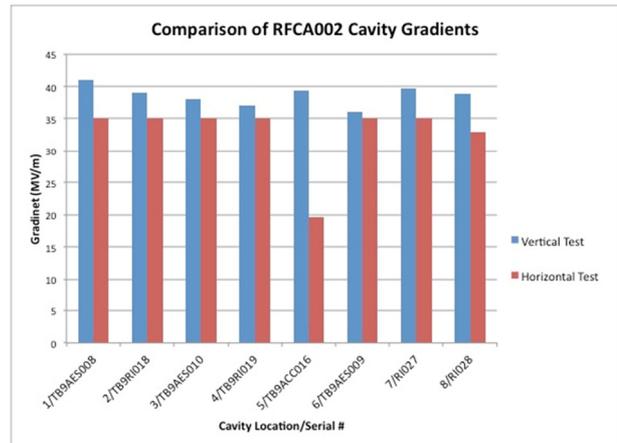

Figure 2: CM-2 peak cavity gradients as measured vertically (blue) and at Fermilab's HTS (red).

## SUMMARY

Testing of the first Tesla-style SRF module at Fermilab has been successfully completed and preparation for installing and bringing a second module, capable of meeting ILC specifications, is well underway. With these achievements as well as operation of other SRF facilities, Fermilab is demonstrating its maturity in this field.

## ACKNOWLEDGEMENTS

The achievements to date would not have been possible without the hard work and effort of many people in Fermilab's Technical and Accelerator Divisions. Shared expertise and contributions of hardware etc. from collaborators from many institutions around the world especially Argonne National Laboratory, DESY, INFN/LASA, Milano, Jefferson Lab, and SLAC, are noted and appreciated.

## REFERENCES

[1] J. Leibfritz et al., 'Status and Plans for a Superconducting RF Accelerator Test Facility at Fermilab," IPAC2012, New Orleans, May 2012, MOOAC02.

[2] E.R. Harms et al., "RF Test Results from Cryomodule 1 at the Fermilab SRF Test Beam Facility," SRF'11, Chicago, July 2011, MOPO013.

[3] Y. Pischalnikov et al., "Lorentz Force Compensation for Long Pulses in SRF Cavities," IPAC2012, New Orleans, May 2012, THPPR012.

[4] A. Hocker et al., "Individual RF Test Results of the Cavities Used in the First US-built ILC-type Cryomodule," IPAC2012, New Orleans, WEPPC049.